# A spintronic analog of the Landauer residual resistivity dipole on the surface of a disordered topological insulator

Raisa Fabiha and Supriyo Bandyopadhyay[1]
*Department of Electrical and Computer Engineering,*
*Virginia Commonwealth University, Richmond, 23284, Virginia, United States*

The Landauer "residual resistivity dipole" is a well-known concept in electron transport through a disordered medium. It is formed when a defect/scatterer reflects an impinging electron causing negative charges to build up on one side of the scatterer and positive charges on the other. This results in the formation of a microscopic electric dipole that affects the resistivity of the medium. Here, we show that an equivalent entity forms in spin polarized electron transport through the surface of a disordered topological insulator (TI). When electrons reflect from a scatterer on the TI surface, a spin imbalance forms around the scatterer, resulting in a spin current that flows either in the same or the opposite direction as the injected spin current and hence either increases or decreases the spin resistivity. It also destroys spin-momentum locking and produces a magnetic field around the scatterer. The latter will cause transiting spins to precess as they pass the scatterer, thereby radiating electromagnetic waves and implementing an oscillator. If an alternating current is passed through the TI instead of a static current, the magnetic field will oscillate with the frequency of the current and radiate electromagnetic waves of the same frequency, thus making the scatterer act as a miniature antenna.

## INTRODUCTION

The Landauer residual resistivity dipole (LRRD) is a familiar concept in microscopic charge transport and has important consequences for electromigration [1, 2]. The basic idea behind the LRRD is illustrated in Fig. 1. A moving electron in a charge current (sometimes referred to as an "electron wind") encounters a scatterer on the way and is reflected with some probability, causing negative charges to accumulate on the impinging side and deplete on the opposite side. This charge imbalance causes an electric dipole to form around the scatterer.

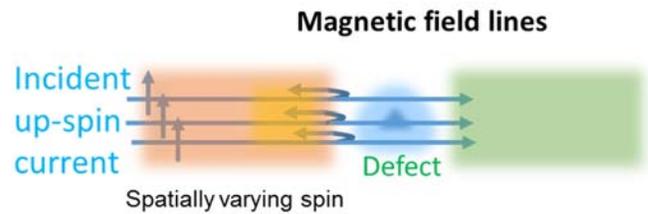

FIG. 2. Formation of a residual magnetic dipole around a scatterer on the surface of a topological insulator due to reflection of spins. (a) Reflection without spin flip, and (b) reflection with spin flip. Here we have assumed that the transmission occurs without a spin flip.

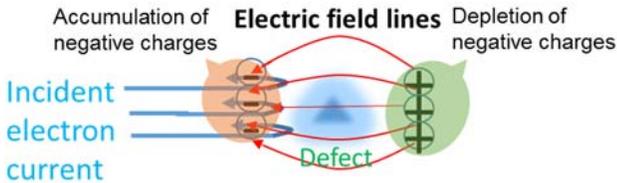

FIG. 1. Electrons in a current reflect from a defect, causing negative charges to pile up on the impinging side and positive charges on the other side. This charge imbalance gives rise to a microscopic electric dipole around the impurity, known as the Landauer residual resistivity dipole.

It is natural to ask if an equivalent *magnetic* entity can exist in *spin polarized* electron transport. We show that a similar phenomenon can occur on the surface of a disordered topological insulator (TI) through which a spin polarized current is flowing. As the impinging spins reflect from a static scatterer, with or without a spin flip, the spin polarizations on both sides of the scatterer vary spatially owing to interference between the incident and reflected waves, causing a spin imbalance to form between the two sides. This then causes a spin current to flow either in the same or in the opposite direction of the injected current, depending on the sign of the imbalance. These scattering-induced spin currents obviously aid or oppose the injected spin current, thereby increasing or decreasing the "spin resistivity". This is depicted schematically in Fig. 2. The spin imbalance also forms a local magnetic field around the scatterer, which can cause transiting spins to precess and radiate electromagnetic waves, making the scatterer act as a source of radiation. Here, we analyze this phenomenon.

## THEORY

For the sake of simplicity, we will consider a *line defect* on the surface of the TI (perpendicular to the direction of current flow). It can be created by implanting a row of magnetic impurities using ion implantation. A magnetic impurity can reflect a spin with or without spin flip. Fig. 3 shows such a system. In the figure we depict the case when the impinging spin reflects and transmits *without* a spin flip, but in the ensuing theory, we account for reflection/transmission both with and without any spin flip.

To make the mathematics simple, we will consider a line defect of zero spatial width on the surface of a topological insulator (TI).

[1] Email: sbandy@vcu.edu

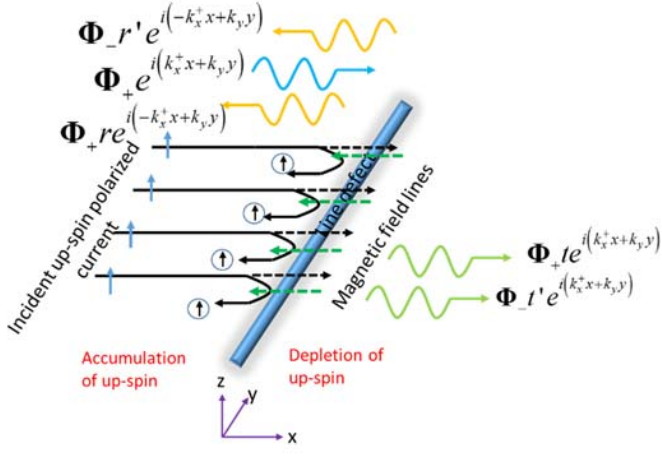

FIG. 3. Reflection of a spin polarized current from a line defect on the surface of a topological insulator.

The Pauli equation for the spinor wave function $\psi$ of an electron on the TI surface containing the line defect is [3]

$$-\frac{\hbar^2}{2m^*}\frac{d^2\psi(x,y)}{dx^2} - \frac{\hbar^2}{2m^*}\frac{d^2\psi(x,y)}{dy^2} + \Gamma\delta(x)\psi(x,y)$$
$$+\Gamma'\sigma_f\delta(x)\psi(x,y) - i\hbar v_0\left(\frac{d}{dx}\sigma_y - \frac{d}{dy}\sigma_x\right)\psi(x,y) \quad (1)$$
$$= E\psi(x,y)$$

where the line defect is viewed as a one-dimensional delta-scatterer that has a spin-independent part of strength $\Gamma$ and a spin-dependent part of strength $\Gamma'$. The quantity $v_0$ is the Fermi velocity and $\sigma_f$ is the spin-flip matrix.

Integrating both sides of this equation from $-\varepsilon$ to $+\varepsilon$ and letting $\varepsilon \to 0$, we get [4]

$$\left.\frac{d\psi(x,y)}{dx}\right|_{x=0+} - \left.\frac{d\psi(x,y)}{dx}\right|_{x=0-}$$
$$= \frac{2m^*}{\hbar^2}\Gamma\left[1+\left(\frac{\Gamma'}{\Gamma}\sigma_f\right)\right]\psi(x=0,y) \quad (2)$$

In deriving the above equation, we made use of the continuity of the wave function at $x = 0$.

Note that since the Hamiltonian in Equation (1) is invariant in the coordinate $y$, the $y$-component of the wave vector $k_y$ is a good quantum number. However, the $x$-components of the wave vectors in the two eigenspinor states on the TI surface are different for any given energy [5] and these two wave vectors will be denoted as $k_x^\pm$.

We can write the wave functions in the two spin eigenstates on a pristine TI surface (without any defect) as

$$\psi_\pm(x,y) = \Phi_\pm e^{i(k_x^\pm x + k_y y)}, \quad (3)$$

where the eigenspinors are given by [5]

$$\Phi_\pm = \frac{1}{\sqrt{2}}\begin{bmatrix} 1 \\ \mu_\pm \end{bmatrix},$$
$$\mu_\pm = \pm i(k_x^\pm/k^\pm) \mp (k_y/k^\pm) \quad (4)$$

and $k^\pm = \sqrt{(k_x^\pm)^2 + k_y^2}$.

We will next consider the situation where an electron in the $\Phi_+$ eigenstate is incident on the line defect at $x = 0$. The other case, when the incident electron is in the $\Phi_-$ eigenstate, is similar and is omitted here for the sake of brevity. Let the reflection amplitude for reflecting without a spin flip be $r$, with a spin flip be $r'$ and the transmission amplitudes without and with a spin flip be $t$ and $t'$, respectively. The wavefunction to the left of the line defect is

$$\psi_L(x,y) = \Phi_+ e^{i(k_x^+ x + k_y y)} + r\Phi_+ e^{i(-k_x^+ x + k_y y)} + r' e^{i(-k_x^+ x + k_y y)}\Phi_-$$

while to the right, it is $\psi_R(x,y) = t\Phi_+ e^{i(k_x^+ x + k_y y)} + t'\Phi_- e^{i(k_x^+ x + k_y y)}$ [see Fig. 3].

Enforcing the continuity of the wave function at $x = 0$, we get

$$[1+r]\Phi_+ + r'\Phi_- = t\Phi_+ + t'\Phi_- \quad (5)$$

Next, using Equation (3) in (2), we obtain

$$ik_x^+[1-r]\Phi_+ - ik_x^- r'\Phi_- - ik_x^+ t\Phi_+ - ik_x^- t'\Phi_-$$
$$= -\frac{2m^*}{\hbar^2}\Gamma\left[(1+r)\Phi_+ + r'\Phi_-\right]$$
$$-\frac{2m^*}{\hbar^2}\Gamma'\sigma_f\left[(1+r)\Phi_+ + r'\Phi_-\right]$$
$$= -\frac{2m^*}{\hbar^2}\Gamma\left[(1+r)\Phi_+ + r'\Phi_-\right]$$
$$-\frac{2m^*}{\hbar^2}\Gamma'\left[(1+r)\Phi_- + r'\Phi_+\right]. \quad (6)$$

With the aid of Equation (4), Equation (5) can be written as

$$[\mathbf{A}]\left(\begin{bmatrix} t \\ t' \end{bmatrix} - \begin{bmatrix} r \\ r' \end{bmatrix}\right) = [\mathbf{C}],$$

where $\quad (7)$

$$[\mathbf{A}] = [\Phi_+ \quad \Phi_-] = \frac{1}{\sqrt{2}}\begin{bmatrix} 1 & 1 \\ \mu_+ & \mu_- \end{bmatrix}$$
$$[\mathbf{C}] = \Phi_+ = \frac{1}{\sqrt{2}}\begin{bmatrix} 1 \\ \mu_+ \end{bmatrix}$$

Note that the matrix $[\mathbf{A}]$ is not unitary since $k_x^+ \neq k_x^-$.

From Equation (7), we get

$$\begin{bmatrix} t \\ t' \end{bmatrix} = \begin{bmatrix} r \\ r' \end{bmatrix} + [\mathbf{A}]^{-1}[\mathbf{C}]. \tag{8}$$

Then from Equation (6), we obtain

$$\left(ik_x^+ + \frac{2m^*}{\hbar^2}\Gamma\right)\Phi_+ + \frac{2m^*}{\hbar^2}\Gamma'\Phi_-$$

$$= r\left[\left(ik_x^+ - \frac{2m^*}{\hbar^2}\Gamma\right)\Phi_+ - \frac{2m^*}{\hbar^2}\Gamma'\Phi_-\right], \tag{9}$$

$$+ r'\left[\left(ik_x^- - \frac{2m^*}{\hbar^2}\Gamma\right)\Phi_- - \frac{2m^*}{\hbar^2}\Gamma'\Phi_+\right]$$

$$+ t ik_x^+ \Phi_+ + t' ik_x^+ \Phi_-$$

which can be written in matrix form as

$$[\mathbf{B}]\begin{bmatrix} r \\ r' \end{bmatrix} + [\mathbf{D}]\begin{bmatrix} t \\ t' \end{bmatrix} = [\mathbf{K}], \tag{10}$$

where

$$[\mathbf{B}] = [\mathbf{b}_1 \quad \mathbf{b}_2]$$

$$\mathbf{b}_1 = \left(ik_x^+ - \frac{2m^*}{\hbar^2}\Gamma\right)\Phi_+ - \frac{2m^*}{\hbar^2}\Gamma'\Phi_-$$

$$\mathbf{b}_2 = \left(ik_x^- - \frac{2m^*}{\hbar^2}\Gamma\right)\Phi_- - \frac{2m^*}{\hbar^2}\Gamma'\Phi_+$$

$$\therefore [\mathbf{B}] = \frac{1}{\sqrt{2}}\begin{bmatrix} b_{11} & b_{12} \\ b_{21} & b_{22} \end{bmatrix} \text{ where}$$

$$b_{11} = ik_x^+ - \frac{2m^*}{\hbar^2}(\Gamma + \Gamma'); b_{12} = ik_x^- - \frac{2m^*}{\hbar^2}(\Gamma + \Gamma')$$

$$b_{21} = \left(ik_x^+ - \frac{2m^*}{\hbar^2}\Gamma\right)\mu_+ - \frac{2m^*}{\hbar^2}\Gamma'\mu_-$$

$$b_{22} = \left(ik_x^- - \frac{2m^*}{\hbar^2}\Gamma\right)\mu_- - \frac{2m^*}{\hbar^2}\Gamma'\mu_+$$

$$[\mathbf{D}] = [ik_x^+ \Phi_+ \quad ik_x^- \Phi_-]$$

$$= \frac{1}{\sqrt{2}}\begin{bmatrix} ik_x^+ & ik_x^- \\ (ik_x^+)\mu_+ & (ik_x^-)\mu_- \end{bmatrix}.$$

$$[\mathbf{K}] = \left(ik_x^+ + \frac{2m^*}{\hbar^2}\Gamma\right)\Phi_+ + \frac{2m^*}{\hbar^2}\Gamma'\Phi_-$$

$$= \frac{1}{\sqrt{2}}\left(ik_x^+ + \frac{2m^*}{\hbar^2}\Gamma\right)\begin{bmatrix} 1 \\ \mu_+ \end{bmatrix} + \frac{1}{\sqrt{2}}\left(\frac{2m^*}{\hbar^2}\Gamma'\right)\begin{bmatrix} 1 \\ \mu_- \end{bmatrix}$$

Using Equation (8) in Equation (10), we obtain a solution for the reflection amplitudes as

$$\begin{bmatrix} r \\ r' \end{bmatrix} = ([\mathbf{B}]+[\mathbf{D}])^{-1}[\mathbf{K}] \tag{11}$$

$$-([\mathbf{B}]+[\mathbf{D}])^{-1}[\mathbf{D}][\mathbf{A}]^{-1}[\mathbf{C}]$$

Finally, using Equation (11) in Equation (8), we get the solution for the transmission amplitudes:

$$\begin{bmatrix} t \\ t' \end{bmatrix} = ([\mathbf{B}]+[\mathbf{D}])^{-1}[\mathbf{K}] - \{([\mathbf{B}]+[\mathbf{D}])^{-1}[\mathbf{D}]-[\mathbf{I}]\}[\mathbf{A}]^{-1}[\mathbf{C}], \tag{12}$$

where $[\mathbf{I}]$ is the $2\times 2$ identity matrix.

The wave function on the left of the line defect is (see Fig. 3)

$$\psi_L = \Phi_+ e^{i(k_x^+ x + k_y y)} + r\Phi_+ e^{i(-k_x^+ x + k_y y)}$$

$$+ r'\Phi_- e^{i(-k_x^- x + k_y y)}$$

$$= \frac{1}{\sqrt{2}}\begin{bmatrix} 1 \\ \mu_+ \end{bmatrix} e^{i(k_x^+ x + k_y y)}$$

$$+ r\frac{1}{\sqrt{2}}\begin{bmatrix} 1 \\ \mu_+ \end{bmatrix} e^{i(-k_x^+ x + k_y y)} \tag{13}$$

$$+ r'\frac{1}{\sqrt{2}}\begin{bmatrix} 1 \\ \mu_- \end{bmatrix} e^{i(-k_x^- x + k_y y)}$$

$$= \begin{bmatrix} \alpha_1(x,y) \\ \alpha_2(x,y) \end{bmatrix}$$

whereas on the right it is

$$\psi_R = t\Phi_+ e^{i(k_x^+ x + k_y y)} + t'\Phi_- e^{i(k_x^- x + k_y y)}$$

$$= t\frac{1}{\sqrt{2}}\begin{bmatrix} 1 \\ \mu_+ \end{bmatrix} e^{i(k_x^+ x + k_y y)}$$

$$+ t'\frac{1}{\sqrt{2}}\begin{bmatrix} 1 \\ \mu_- \end{bmatrix} e^{i(k_x^- x + k_y y)} \tag{14}$$

$$= \begin{bmatrix} \beta_1(x,y) \\ \beta_2(x,y) \end{bmatrix}$$

Therefore, the $x$-, $y$- and $z$-components of the spin on the left of the line defect are

$$S_x^L(x,y) = \hbar \operatorname{Re}[\alpha_1^*(x,y)\alpha_2(x,y)]/M(x,y)$$
$$S_y^L(x,y) = \hbar \operatorname{Im}[\alpha_1^*(x,y)\alpha_2(x,y)]/M(x,y) \tag{15}$$
$$S_z^L(x,y) = \frac{\hbar}{2}[|\alpha_1|^2(x,y) - |\alpha_2|^2(x,y)]/M(x,y)$$

and the same components on the right of the defect are

$$S_x^R(x,y) = \hbar \operatorname{Re}[\beta_1^*(x,y)\beta_2(x,y)]/M'(x,y)$$
$$S_y^R(x,y) = \hbar \operatorname{Im}[\beta_1^*(x,y)\beta_2(x,y)]/M'(x,y) \tag{16}$$
$$S_z^R(x,y) = \frac{\hbar}{2}[|\beta_1|^2(x,y) - |\beta_2|^2(x,y)]/M'(x,y)$$

where $M(x,y) = \sqrt{|\alpha_1(x,y)|^2 + |\alpha_2(x,y)|^2}$ and $M'(x,y) = \sqrt{|\beta_1(x,y)|^2 + |\beta_2(x,y)|^2}$.

# RESULTS

We assume $m^* = 0.1\ m_0$ (free electron mass), $(1/2)m_0 v_0^2 = 100$ meV, $\Gamma = 6\times 10^{-29}$ J-m and $\Gamma' = 4\times 10^{-29}$ J-m. The energy dispersion relation on the surface of a topological insulator (without any defect) is given by [5]

$$E = \frac{\hbar^2(k_x^2 + k_y^2)}{2m^*} \pm \hbar v_0 \sqrt{k_x^2 + k_y^2}. \quad (17)$$

For the sake of simplicity, we will consider the case for $k_y = 0$, in which case the $x$-components of the wavevectors in the two spin eigenstates are related to the energy $E$ as

$$k_x^+ = \frac{m^* v_0}{\hbar} + \sqrt{\left(\frac{m^* v_0}{\hbar}\right)^2 + \frac{2m^* E}{\hbar^2}}$$
$$k_x^- = -\frac{m^* v_0}{\hbar} + \sqrt{\left(\frac{m^* v_0}{\hbar}\right)^2 + \frac{2m^* E}{\hbar^2}} \quad (18)$$

Note that when $k_y = 0$, the incident spin is completely $y$-polarized because of spin-momentum locking on a TI surface.

We use Equation (18) in Equations (11) and (12) to find the transmission and reflection probabilities as functions of the electron energy $E$ for $k_y = 0$. They are plotted in Fig. 4. We have verified that the current continuity condition is always satisfied at every energy, i.e.

$$|t(E)|^2 + |r(E)|^2 + \frac{k_x^-}{k_x^+}|t'(E)|^2 + \frac{k_x^-}{k_x^+}|r'(E)|^2 = 1. \quad (19)$$

For $k_y = 0$, we find that $[\mathbf{A}]^{-1}[\mathbf{C}] = [1\ \ 0]\dagger$ and hence from Equation (8), $t = 1 + r$ and $t' = r'$.

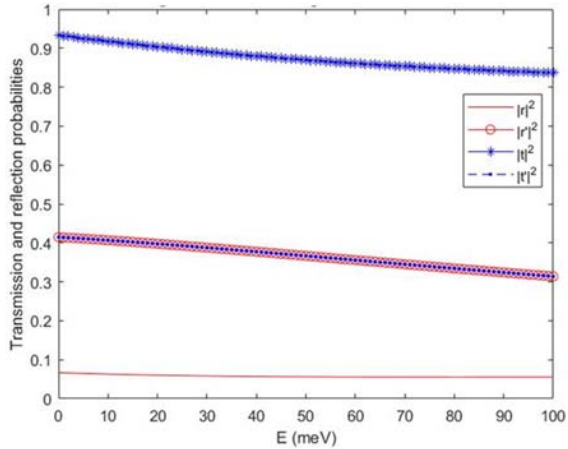

FIG. 4. Transmission and reflection probabilities as functions of electron energy for $\Gamma = 6\times 10^{-29}$ J-m and $\Gamma' = 4\times 10^{-29}$ J-m, and $k_y = 0$.

For $k_y = 0$, $\mu_\pm = \pm i$, and hence we get from Equations (13) and (14):

$$\alpha_1(x) = \frac{1}{\sqrt{2}}\left(e^{ik_x^+ x} + re^{-ik_x^+ x} + r'e^{-ik_x^- x}\right)$$
$$\alpha_2(x) = \frac{i}{\sqrt{2}}\left(e^{ik_x^+ x} + re^{-ik_x^+ x} - r'e^{-ik_x^- x}\right)$$
$$\beta_1(x) = \frac{1}{\sqrt{2}}\left(te^{ik_x^+ x} + t'e^{ik_x^- x}\right) \quad (20)$$
$$\beta_2(x) = \frac{i}{\sqrt{2}}\left(te^{ik_x^+ x} - t'e^{ik_x^- x}\right)$$

We use these relations in Equations (15) and (16) to find the spin components as functions of the distances from the scatterer on both sides extending up to 25 nm for an energy $E = 100$ meV and plot them in Fig. 5.

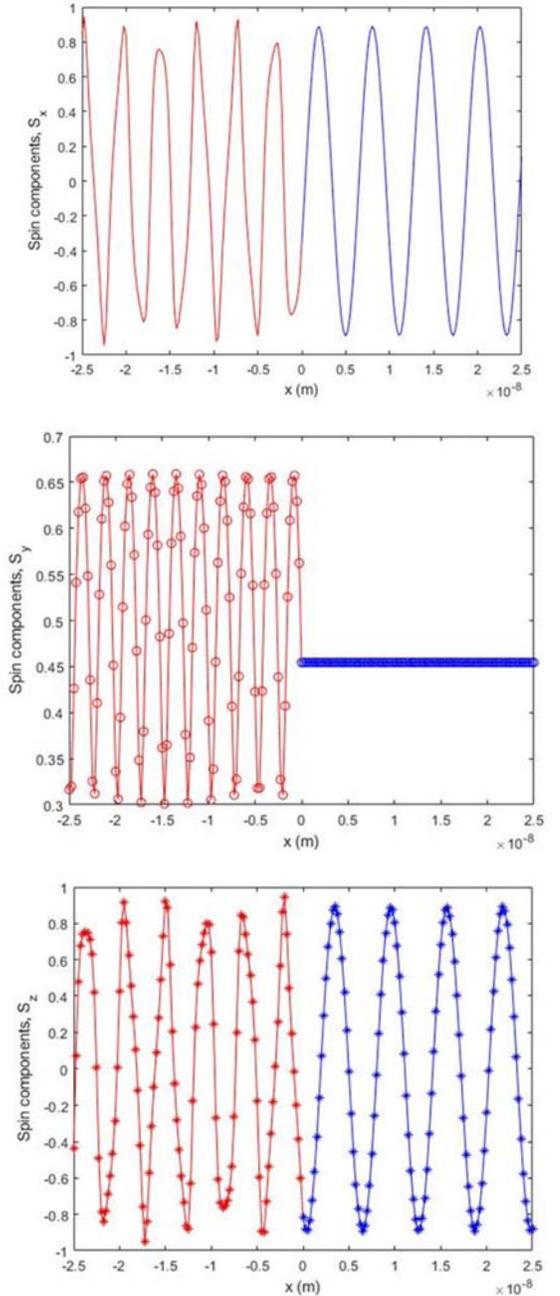

FIG. 5. Spin components to the left and right of the line defect, as functions of the distance from the scatterer with $k_y = 0$.

We see from Fig. 5 that the spin components to the left and right of the line defect are *not* the same at any given distance from the scatterer. Since the incident spin is *y*-polarized, the *y*-component of the transmitted spin (and hence the *y*-component on the right hand side) is spatially invariant, while the other components oscillate in space in the manner of a spin helix owing to interference between the incident and reflected spin. Clearly, spin-momentum locking is destroyed. We define spatially averaged spin components on the two sides of the scatterer as

$$\langle S_i^L \rangle = \int_{-W}^{0} S_i^L(x)\,dx \quad [i = x, y, z],$$

and $\langle S_i^R \rangle = \int_{0}^{W} S_i^R(x)\,dx \quad [i = x, y, z]$, and then list them in Table 1 in arbitrary units for an arbitrary value of *W* = 25 nm.

**Table 1: Average spin over a fixed distance *W* on the two sides of the scatterer with *W* = 25 nm.**

| $\langle S_x^L \rangle$ | $\langle S_x^R \rangle$ | $\langle S_y^L \rangle$ | $\langle S_y^R \rangle$ | $\langle S_z^L \rangle$ | $\langle S_z^R \rangle$ |
|---|---|---|---|---|---|
| -0.011 | -0.003 | 0.517 | 0.459 | 0.021 | -0.027 |

Table 1 shows that there is a net spin imbalance between the two sides of the line defect as we move away from the defect by any arbitrary distance on both sides, causing the formation of a local magnetization and an associated magnetic field. The spin imbalance will also cause a spin current to flow which can aid or oppose the injected spin current depending on the sign of the imbalance, thereby decreasing or increasing the spin resistivity. It is therefore a spintronic analog of the LRRD.

In Fig. 6, we plot the angular separation between the spin polarizations on the two sides of the scatterer as a function of the distance from the scatterer. This quantity $\theta(x)$ is defined as

$$\cos\theta(x) = \frac{S_x^L(-x)S_x^R(x) + S_y^L(-x)S_y^R(x) + S_z^L(-x)S_z^R(x)}{\sqrt{\left[S_x^L(-x)\right]^2 + \left[S_y^L(-x)\right]^2 + \left[S_z^L(-x)\right]^2}\sqrt{\left[S_x^R(x)\right]^2 + \left[S_y^R(x)\right]^2 + \left[S_z^R(x)\right]^2}}$$

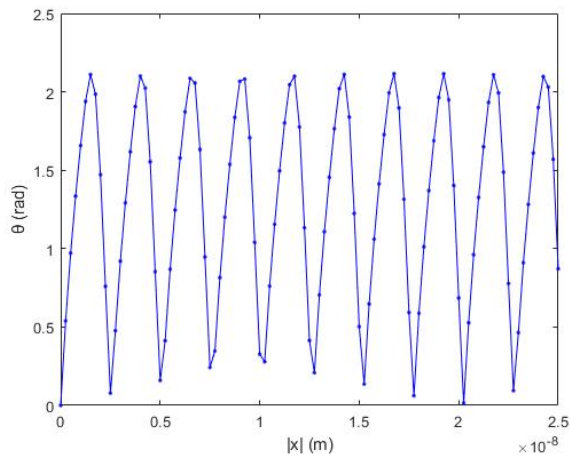

FIG. 6. Angular separation between the spin polarizations on two sides of the scatterer as a function of the distance |*x*| from the scatterer.

Because the spin components oscillate in space, the angular separation oscillates as well and the maximum angular separation exceeds $120^0$ for this case.

**CONCLUSION**

In this work, we have shown the existence of a spin imbalance around a line defect on the surface of a current-carrying topological insulator, reminiscent of the LRRD that causes a charge imbalance. Its existence can be verified experimentally with magnetic force microscopy.

The magnetic field resulting from the spin imbalance will make spins transiting through the defect precess and radiate electromagnetic waves [6-9], thereby making the defect act as an electromagnetic *oscillator* or *radiator*, provided the damping is relatively small. The sample will be a broadband oscillator since the precession frequencies will be different at different scattering sites. Finally, if instead of a static current, we use an alternating current, then the local magnetic fields will oscillate in time with the same frequency as the injected current. That too can radiate electromagnetic waves, making the line defect act as a miniature *antenna* [10-13]. These radiations can be detected with suitable detectors.

**REFERENCES**

1. R. Landauer, "Residual resistivity dipoles", *Zeitschrift für Physik B Condensed Matter*, **21**, 247-254 (1975).
2. W. Zwerger, L. Bönig and K. Schönhammer, "Exact scattering theory for the Landauer residual resistivity dipole", *Phys. Rev. B*, **43**, 6434-6439 (1991).
3. W. Jung, et al., "Warping effects in the band and angular momentum structures of the topological insulator $Bi_2Te_3$", *Phys. Rev. B*, **84,** 245435 (2011).
4. M. Cahay and S. Bandyopadhyay, *Problem Solving in Quantum Mechanics* (Wiley, Chichester, UK, 2017).
5. S. Shee, R. Fabiha, M. Cahay and S. Bandyopadhyay, "Reflection and refraction of a spin at the edge of a quasi-two-dimensional semiconductor layer (quantum well) and a topological insulator", *Magnetism* (in press).
6. S. Bandyopadhyay and M. Cahay, *Introduction to Spintronics*, 2nd edition (CRC Press, Boca Raton, 2015).
7. D. J. Griffiths, *Introduction to Electrodynamics*, 4th ed. (Pearson, London, 2013)
8. R. Rungsawang, et al., "Terahertz radiation from magnetic excitations in diluted magnetic semiconductors", *Phys. Rev. Lett.*, **110**, 177203 (2013).
9. O. Dzyapko, V. E. Demidov, S. O. Demokritov, G. A. Melkov and V. L. Safanov, "Monochromatic microwave radiation from the system of strongly excited magnons", *Appl. Phys. Lett.*, **92**, 162510 (2008).
10. J. L. Drobitch, A. De, K. Dutta, P. K. Pal. A. Adhikari, A. Barman and S. Bandyopadhyay, "Extreme sub-wavelength magneto-elastic electromagnetic antenna implemented with multiferroic nanomagnets", *Adv. Mater. Technol.*, **5**, 2000316 (2020).
11. J. D. Schneider, et al., "Experimental demonstration and operating principles of a multiferroic antenna", *J. Appl. Phys.*, **126**, 224104 (2019).
12. S. Prasad M N, Y. Huang and Y. E. Wang, "Going beyond Chu-Harrington limit", ULF radiation with a spinning magnet array", Proc. 32nd URSI GASS, Montreal, August 2017.
13. R. Fabiha, J. Lundquist, S. Majumder, E. Topsakal, A. Barman and S. Bandyopadhyay, "Spin wave electromagnetic nano-antenna enabled by tripartite phonon-magnon-photon coupling, *Adv. Sci.*, 2104644 (2022).